\documentclass[amsmath,amssymb,aps,prl,twocolumn,showpacs,superscriptaddress]{revtex4-1}
\usepackage{graphicx}
\usepackage{dcolumn}
\usepackage{bm}
\usepackage{xr}
\makeatletter
\usepackage{hyperref}
\usepackage{xcolor}
\hypersetup{
  colorlinks   = true,
  urlcolor     = black,
  linkcolor    = blue,
  citecolor   = blue
}

\begin{document}

\title{Diffusion through nanopores in connected lipid bilayer networks}

\author{M. Valet}
\affiliation{Laboratoire Jean Perrin CNRS UMR 8237, Sorbonne Universit\'e, 4 place Jussieu, 75005 Paris, France}
\author{L.-L. Pontani}
\affiliation{Laboratoire Jean Perrin CNRS UMR 8237, Sorbonne Universit\'e, 4 place Jussieu, 75005 Paris, France}
\author{R. Voituriez}
\affiliation{Laboratoire Jean Perrin CNRS UMR 8237, Sorbonne Universit\'e, 4 place Jussieu, 75005 Paris, France}
\affiliation{Laboratoire de Physique Th\'eorique de la Mati\`ere Condens\'ee, CNRS UMR 7600, Sorbonne Universit\'e, 4 place Jussieu, 75005 Paris, France}
\author{E. Wandersman}
\email[]{elie.wandersman@sorbonne-universite.fr}
\affiliation{Laboratoire Jean Perrin CNRS UMR 8237, Sorbonne Universit\'e, 4 place Jussieu, 75005 Paris, France}
\author{A. M. Prevost}
\email[]{alexis.prevost@sorbonne-universite.fr}
\affiliation{Laboratoire Jean Perrin CNRS UMR 8237, Sorbonne Universit\'e, 4 place Jussieu, 75005 Paris, France}

\date{\today}

\begin{abstract}
A biomimetic model of cell-cell communication was developed to probe the passive molecular transport across ion channels inserted in synthetic lipid bilayers formed between contacting droplets arranged in a linear array. Diffusion of a fluorescent probe across the array was measured for different pore concentrations. The diffusion characteristic time scale is found to vary non-linearly with the pore concentration. Our measurements are successfully modeled by a continuous time random walk description, whose waiting time is the first exit time from a droplet through a cluster of pores. The size of the cluster of pores is found to increase with their concentration. Our results provide a direct link between the mesoscopic permeation properties and the microscopic characteristics of the pores such as their number, size and spatial arrangement.
\end{abstract}

%\pacs{05.40.Fb,87.16.dp}

\maketitle

In multicellular organisms, cell-cell communication is essential for morphogenesis, cell growth and differentiation as well as cell homeostasis~\cite{Bloemendal2013}. Cells have thus developed various mechanisms to communicate with each other, such as the release of solutes/vesicles in their environment, electrical signals and direct cell-cell contacts. Within direct cell-cell contacts, communication through molecular exchange is made possible with protein gates that create nanopores spanning between apposed cytoplasmic cell membranes. Plants and fungi for instance, use respectively the so-called plamosdesmata and septal pores. In animals, two kinds of pores, gap junction channels and tunneling nanotubes have also been identified and are very similar in their structure and function to their plants and fungi counterparts. Gap junctions in particular, consist of juxtaposed protein based hemichannels that can assemble into clustered structures of typical size a few hundreds of nanometers~\cite{Hervederangeon2013}. They enable a passive diffusion--based transport of small hydrophilic molecules between connected cells, whose properties are mostly measured for cells \textit{in vitro} with dye transfer techniques~\cite{abbaci2008advantages}, such as gap-FRAP~\cite{Wade525} for instance, allowing to determine the permeability of the gap junction.\\
\indent In recent years, the use of well controlled artificial multicellular systems to design complex reaction-diffusion processes within the framework of bottom-up synthetic biology has considerably increased. In particular, synthetic membranes such as Droplet Interface Bilayers (DIBs)~\cite{Leptihn2013}, that are obtained by putting in contact aqueous droplets bathing in an oil-lipid mixture, have allowed the study of molecular transport through both passive ion channels~\cite{Heron2009,walsh2016} and active transporters~\cite{Findlay2016}, using fluorescence imaging. Very recently, networks of DIBs connected by passive staphylococcal $\alpha$-hemolysin pores ($\alpha$HL), were used to probe genetically engineered reaction-diffusion based processes~\cite{Dupin2018}. In all these experiments, both for cells \textit{in vitro} and for artificial systems, diffusion processes across either gap junctions or DIBs decorated with nanopores, are usually modeled with a Fick's law combined with a phenomenological permeation law through a membrane, yielding a large scale effective diffusion coefficient. The microscopic mechanisms that underly the permeation law and therefore control the value of this effective diffusion coefficient are however poorly described. In particular, its dependence with the concentration of the nanopores in the membrane and their spatial arrangement has never been evidenced experimentally, nor modeled theoretically.\\
\indent In this Letter, we report a thorough study of the diffusion of molecular probes through linear networks of aqueous droplets connected by DIBs decorated with $\alpha$HL pores at different concentrations. We model the diffusion from one droplet to its neighbors with a continuous time random walk model whose waiting time is the first exit time from a droplet, either through independent pores or through clusters of pores. Our experimental measurements strongly suggest that the diffusion law is controlled by the clustering of the nanopores, and provide estimates of the clusters size as a function of $\alpha\text{HL}$'s concentration.

\begin{figure}[!h]
\includegraphics[width=0.45\textwidth]{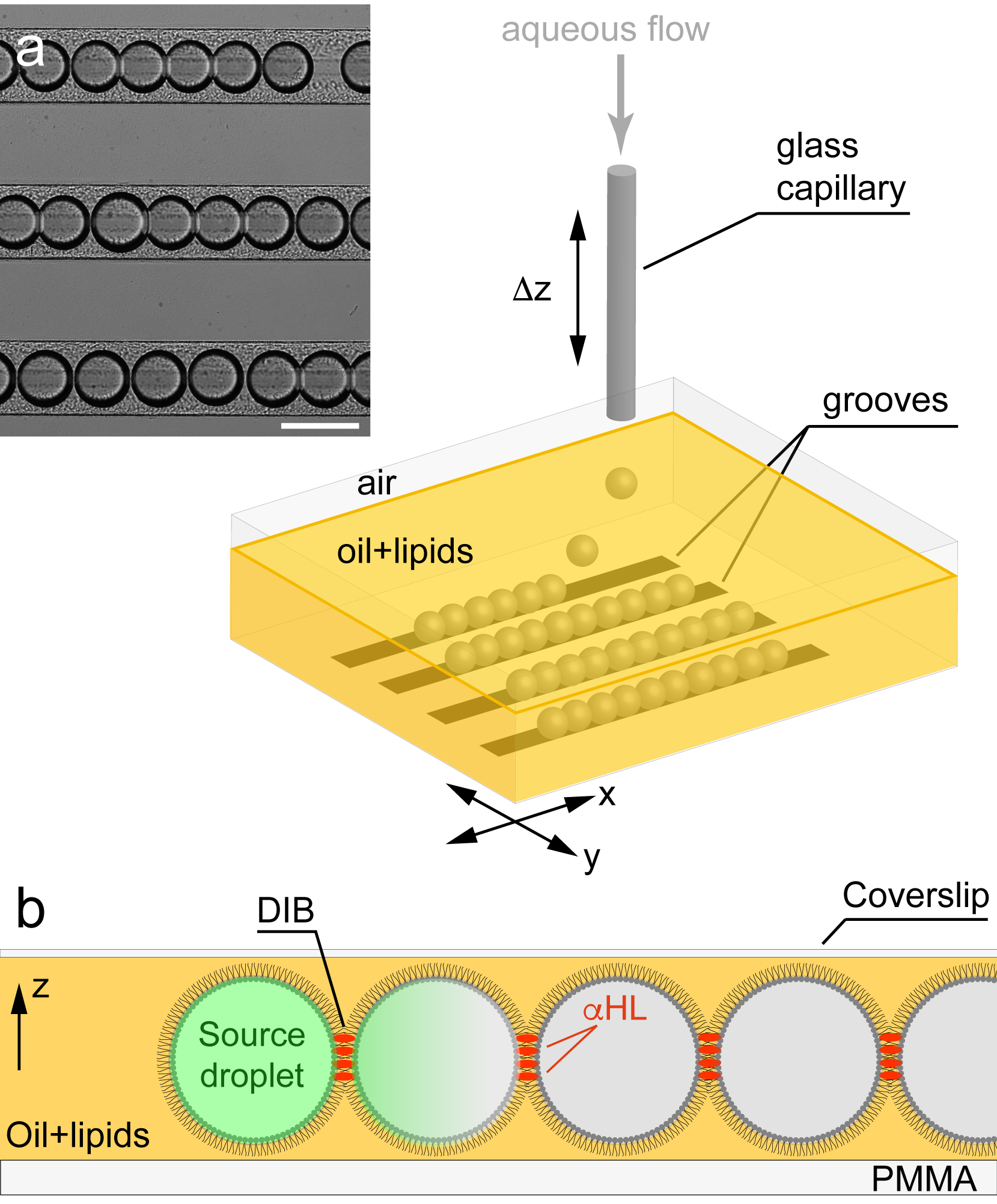}\caption{(a) Sketch of the experimental setup. The aqueous solution is injected through a glass capillary pulled out periodically across the oil/lipid-air interface. The detached droplets sediment in a Plexiglas pool decorated with parallel grooves and form a DIB network. Lateral positioning of the droplets is achieved  by moving the pool with a motorized translation stage. Eventually, a source droplet seeded with fluorophores is deposited. Diffusion of fluorophores is imaged in epifluorescence microscopy. Inset: Image of three typical DIB networks. The white bar is 200~$\mu$m long. (b) Sketch of the diffusion process of fluorophores across DIBs decorated with $\alpha$HL protein pores.  
\label{Fig:Fig1}}
\end{figure}

\par Monomers of $\alpha\text{HL}$ (Sigma Aldrich) were diluted in an aqueous buffer (HEPES $10$~mM, KCl $100$~mM, Sigma Aldrich, $\text{pH}~7.4$) at concentrations $c$ ranging from $150$ to $300~\mu$g/mL. DPhPC lipids in chloroform (4ME 16:0 PC/\textit{1,2-diphytanoyl-sn-glycero-3-phosphocholine}, Avanti) were evaporated under nitrogen and resuspended at a concentration of 6.5~mg/mL in a mixture (50:50 vol:vol) of hexadecane and silicone oil AR20 (Sigma Aldrich).

\begin{figure}
\includegraphics[width=0.45\textwidth]{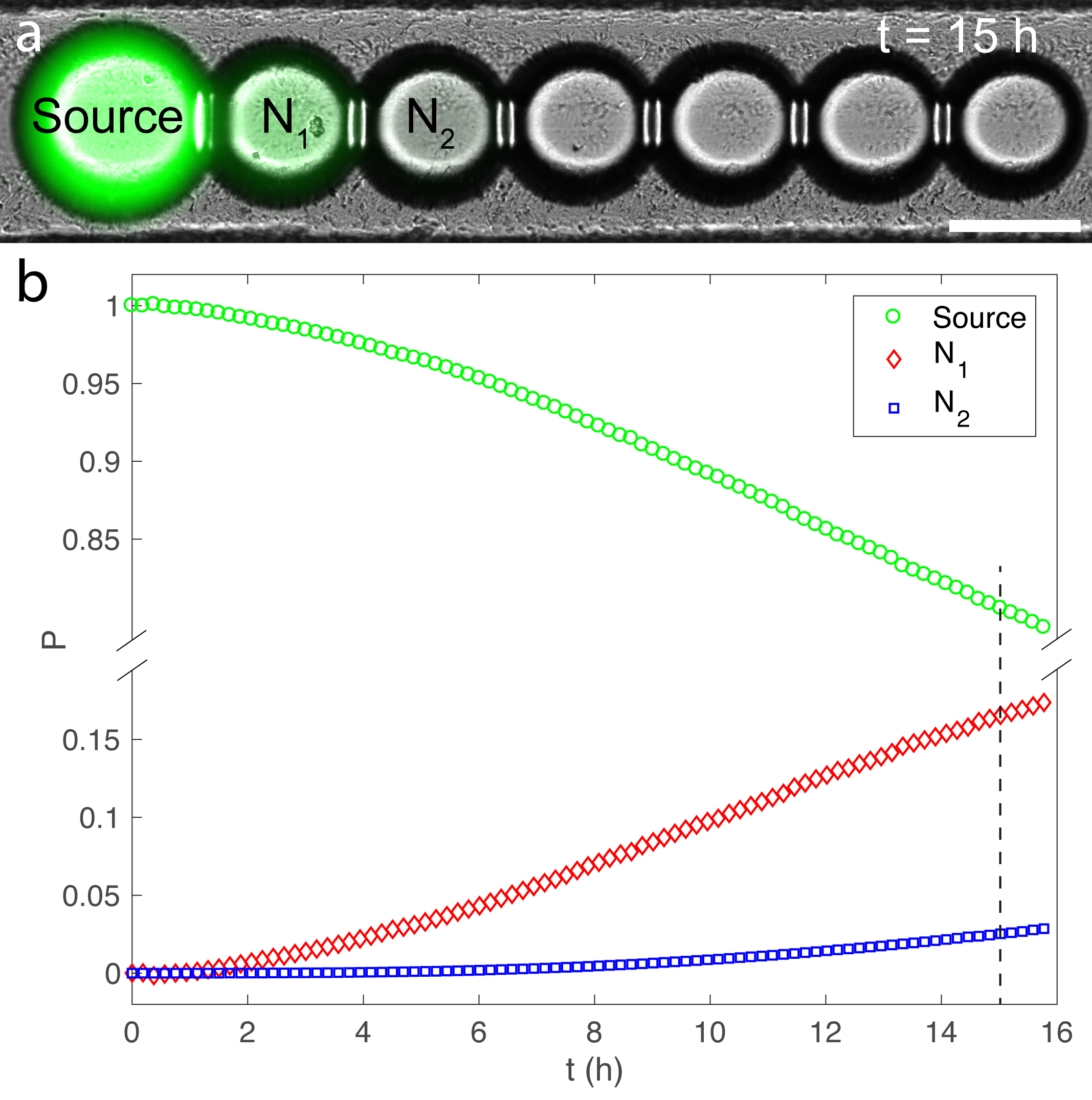}
\caption{(a) Composite image from bright field (grey) and fluorescence (green) microscopy, of a typical diffusion process of carboxyfluorescein across a DIB network ($c=200~\mu$g/mL) at $t=15$~h.  The white bar is 100~$\mu$m long. (b) Occupancy probability $P$ of carboxyfluorescein as a function of time, for the network of (a). From top to bottom, the different curves correspond respectively to the source droplet $S$ and the first $N_1$ and second $N_2$ neighboring droplets. The dashed line corresponds to $t=15$~h.
\label{Fig:Fig2}}
\end{figure}

Droplets of the $\alpha\text{HL}$ solution were produced inside a Plexiglas pool containing the oil/lipid solution and mounted on an XY translation stage, using a Droplet-On-Demand technique~\cite{Valet2018} (Fig.~\ref{Fig:Fig1}a). Briefly, the aqueous phase was flown through a glass capillary (inner diameter $20~\mu$m) periodically extracted through the oil/lipid-air interface. The frequency of this extraction, together with the size of the capillary and the flow rate, control the radius $R$ of the produced droplets. Linear arrays were obtained by depositing droplets of typical $R \approx 75~\mu$m in micromilled grooves (width $200~\mu$m, depth $100~\mu$m) at the bottom of the pool (Fig.~\ref{Fig:Fig1}a). Since each droplet is stabilized with a lipid monolayer, a bilayer was formed at each droplet-droplet contact, thus forming linear arrays of DIBs in which $\alpha\text{HL}$ monomers can heptamerize to form nanopores (Fig.~\ref{Fig:Fig1}b). The adhesion area appearing between neighboring droplets was a signature of a DIB formation (\textit{see} the pairs of white segments at each contact on Fig.~\ref{Fig:Fig2}a).

Source droplets containing a solution of fluorophores (5-carboxyfluorescein from Sigma $20~\mu$M in HEPES 10~mM, KCl~100~mM) were then added either at the end (Fig.~\ref{Fig:Fig2}a) or in the middle of a DIB array. To limit evaporation of the aqueous droplets, a coverslip was placed on the oil/lipid pool (Fig.~\ref{Fig:Fig1}b). Diffusion of fluorophores from the source droplet to its neighbors through the nanopores was imaged in epifluorescence microscopy overnight (typically for 15 hours, every 11~minutes). A typical diffusion process over $16~\text{hours}$ is shown in Fig.~\ref{Fig:Fig2}.

Image analysis was performed on bright field images in order to find the radii and center coordinates of each droplet using a home-made Matlab routine. The fluorescence intensity $I_i(t)$ inside each droplet $N_i$ was measured by taking the average intensity inside a disc that measures half the total droplet size. We checked however that the value of $I_i(t)$ does not change with the disc size. The occupancy probability $P_i$ inside a droplet (except for the source droplet) was then defined as

\begin{equation}
P_i(t)=\frac{I_i(t)-I_i(t_{0})}{I_s(t)+\sum\limits_{j \ne s} (I_j(t)-I_j(t_{0}))}
\label{Eq:Eq1}
\end{equation}

where $I_i(t_{0})$ is the average intensity at the beginning of the experiment, $I_s(t)$ the intensity of the source droplet and where the summation runs over all droplets in the network but the source. For the source droplet, the numerator of Eq.~\ref{Eq:Eq1} is taken as $I_s(t)$, so that $P_s(t_0)=1$. Therefore, $P_s(t)$ decreases with $t$ as fluorophores diffuse from the source to its neighbors, while $P_i(t)$ of the neighboring droplets $N_{i, i \ne s}$ increases (Fig.~\ref{Fig:Fig2}b). Note that a size decrease of the droplets due to evaporation causes an increase of the fluorophore concentration. When evaporation dominates over diffusion, this yields $P_s(t) > 1$ and such cases have thus been excluded from our analysis. In addition, we performed control experiments in the absence of $\alpha$HL nanopores, and measured for 90\% of them no significant increase of $P$ in the network. However, in some cases, an osmotic shock could occur, create transient pores in the DIB, and lead to a rapid increase of the fluorescence signal in immediate neighors over short time scales (typically less than 1~hour). Such fast kinetics events can be easily identified and have also been excluded from the present analysis.

\par
\begin{figure}
\includegraphics[width=0.8\linewidth]{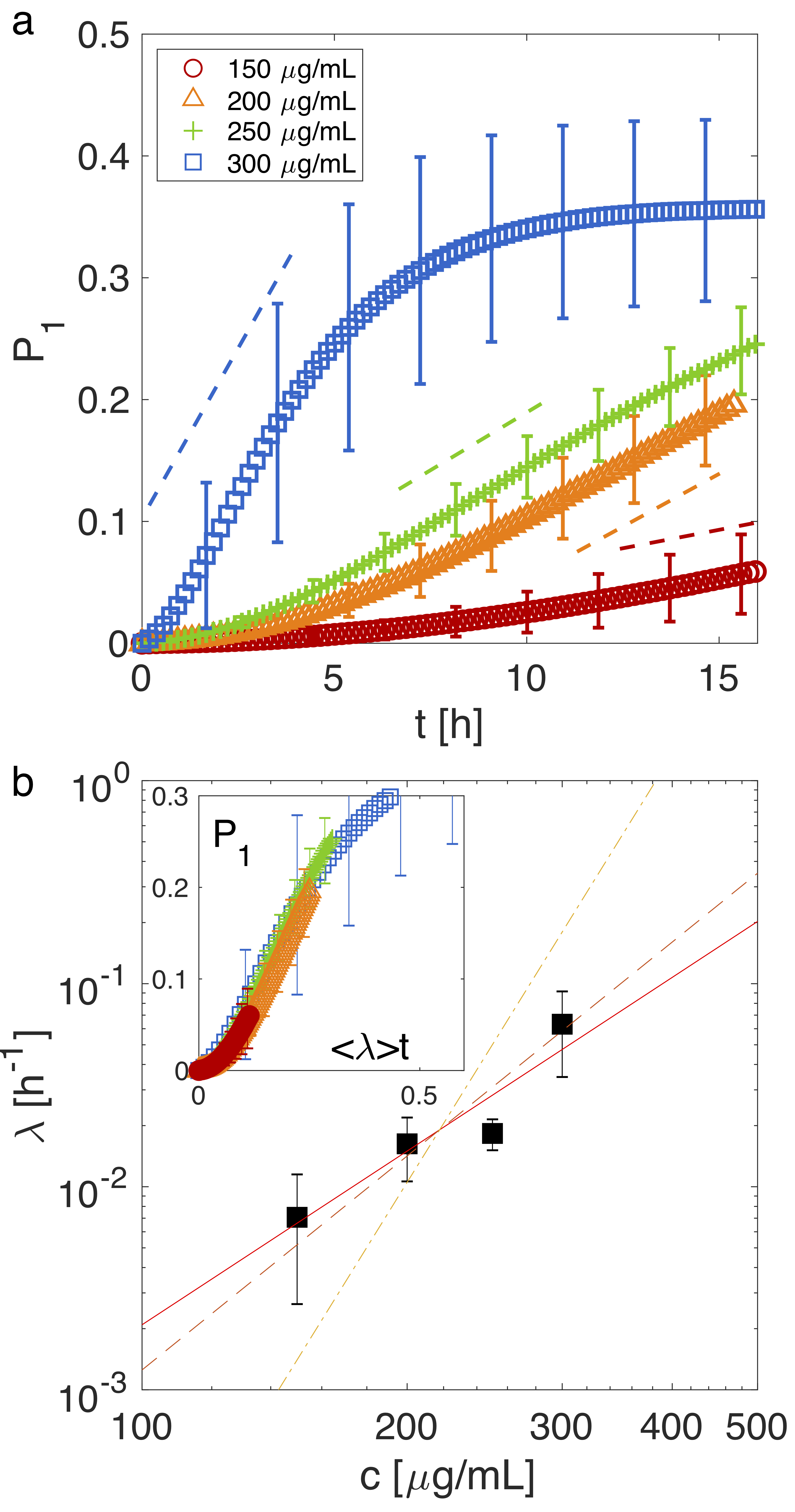}
\caption{(a) Occupancy probability $P_1$ as a function of time $t$ for 4 different $\alpha$HL monomer concentrations, $c$=${150; 200; 250; 300} ~\mu$g/mL, with $N={18; 25; 8; 10}$ respectively. Error bars are standard deviation of the data. The dashed lines are linear fits used to extract the characteristic rate $\lambda$ (\textit{see} main text). They are shifted vertically to ease visualization. (b) Characteristic rate $\lambda$ as a function of $c$. The solid line is a power law fit of exponent 3$\pm$1. The dashed line is a fit using an exponent of 7/2. The dashed dotted line is a fit using and exponent of 7. Inset: Occupancy probability as a function of the rescaled time $<\lambda> t$.
\label{Fig:Fig3}}
\end{figure}

 We quantified the diffusion kinetics by specifically focusing on the source and first neighbor droplets and using different networks with different $\alpha$HL monomer concentrations. The introduced monomers $\alpha$HL$_m$ are first adsorbed and then diffuse within the bilayer to form a heptamer $\alpha$HL. Since the heptamerization process is fast~\cite{Thompson2011}, we describe this chemical sequence as a single step equilibrium $7\,\alpha$HL$_m \leftrightarrow \alpha$HL, which implies that the number of pores adsorbed inside the bilayer scales with the monomer concentration $c$ as $c^7$.\\
Figure~\ref{Fig:Fig3}a shows the occupancy probability $P_1$ for the first neighbor as a function of time $t$ for increasing $\alpha$HL monomer concentrations $c$={150; 200; 250; 300}~$\mu$g/mL. Each curve is an average over several experimental realizations. All curves can be separated within experimental error bars and the diffusion dynamics are faster as the pore concentration increases.
\par Theoretically, for an infinite array of connected compartments, one can model the time evolution of $P_i$ using a continuous time random walk approach~\cite{BookHughes1995}. Each probe molecule is described as a random walker jumping to the adjacent site with a time-dependent probability. The corresponding waiting time is defined as the first exit time of a molecule from a droplet. For 3D brownian diffusion, the latter is known to be exponentially distributed with a mean value that we denote $\tau$~\cite{benichou2010geometry}. Since diffusion within a droplet occurs on time scales much shorter ($\sim 10$~s, with a diffusion coefficient of carboxyfluorescein in water $D=4.10^{-10}m^{2}/s$~\cite{Casalini2011}) than the typical diffusion time from one droplet to its neighbors ($\sim 1$~h), one can assume that the fluorophore concentration is uniform within the droplets, and therefore model the droplets network with a set of discrete connected sites. Within this framework, one can derive using standard tools~\cite{BookHughes1995} the occupancy probability within the first neighbor $P_1^{th}$

\begin{equation}
P_1^{th}(t)=e^{-t/\tau}\mathcal{I}_{1}(t/\tau)
\label{Eq:Besseli}
\end{equation}

\noindent where $\mathcal{I}_{1}$ is the modified Bessel function of the first kind. In our experiments, droplet arrays have a finite size. Thus, Eq.~\ref{Eq:Besseli} cannot model our data at long times. However, in the short time limit ($t/\tau \ll 1$), Eq.~\ref{Eq:Besseli} yields to first order $P^{th} \sim \lambda t$ with $\lambda = 2/\tau$, independently of the total number and position of the droplets.

\par Experimentally, at very short times, our data shows a small deviation from linearity (Fig.~\ref{Fig:Fig3}a) that we interpret as the combined result of both DIBs equilibration and proteins adsorption kinetics at the lipid bilayer. Past this regime, one observes that $P_1(t)$ is approximately linear. For the highest concentration of nanopores, $P_1(t)$ eventually reaches a steady state. For all experimental curves used to compute the average, we have identified this linear regime and chosen to fit it to extract $\lambda$. This was done by testing for the existence of an inflection point at $t=t_{in}$ and extracting the slope $\lambda$ in its vicinity. In case the inflection point was inexistent, we fitted the last 3~hours of the data. Whatever the method used, we checked that the values of $\lambda$ were equally distributed within experimental error bars. For the purpose of illustration, we have performed linear fits on the averaged $P_1(t)$, as shown with the dashed lines on Fig.~\ref{Fig:Fig3}a. We also linearly fitted every single curve to extract $\lambda$ as a function of the nanopore concentration. Resulting $\lambda$'s are shown in Fig.~\ref{Fig:Fig3}b on a log-log plot, and are consistent with a power-law dependence with $c$. Fitting this data with a power law yields $\lambda \sim c^{3\pm1}$. The value of this exponent will be discussed further down theoretically. The inset of Fig.~\ref{Fig:Fig3}b shows the occupancy probability as a function of the rescaled time $<\lambda> t$, where $<\lambda>$ is the average value plotted on the main panel. All curves at different $c$ collapse on the same master curve, indicating that $\lambda^{-1}$ is the only time scale that governs the diffusion kinetics.
\par In recent years, DIBs have increasingly been used ~\cite{booth2017functional,Dupin2018} to study both passive and active transport of molecules using fluorescence based measurements~\cite{Heron2009,walsh2016,Dupin2018}. All these studies assume that diffusion of a molecular probe across a membrane can be described with a phenomenological permeation law, from which an effective large scale permeation coefficient is deduced. However, the microscopic origins that set its value have not been explored. Within the theoretical description presented above (\textit{see} Eq.~\ref{Eq:Besseli}), we propose that the molecular transport from one droplet to another is fully characterized by the mean waiting time $\tau$ in a droplet. This time is the average time necessary for a chemical messenger to reach any single pore of diameter $a$ within the cell, \textit{i.e.} the first exit time~\cite{Schuss2007,Grigoriev2002}. For a spherical domain of radius $R$ and a single pore, this time has been obtained theoretically~\cite{Chevalier2011} and writes

\begin{equation}
<T_1>=\frac{4 \pi R^3}{3 D a}
\end{equation}

\noindent This time is much larger -- by a factor $R/a$ ($\sim 10^5$) -- than the typical time needed for a probe to explore \emph{in bulk} a typical length $R$. For a spherical domain that contains $n$ pores, two different regimes have to be considered. If pores are independent (the typical distance between pores is much larger than $a$), the first passage time $<T_n>$ simply writes $<T_n>=<T_1>/n$. On the contrary, if $n$ pores are clustered, they can be considered as a single pore of area $n a^2$, with a typical size $a^*=\sqrt{n}a$, and thus $<T_n>=<T_1>/\sqrt{n}$. Therefore, how diffusion kinetics depend on the number of nanopores carries information on nanopores spatial organization within the membrane.

Since $\alpha$HL is a heptamer, we expect for both cases the characteristic diffusion rate $\lambda$ to scale either as $c^{7}$ if pores are considered independent, or as $c^{7/2}$ if pores are clustered. Shown in Fig.~\ref{Fig:Fig3}b (dashed line) is a power law fit of $\lambda(c)$ with an exponent $7/2$, in reasonable agreement with the data. On the contrary, a power law with an exponent $7$ is far from providing a quantitative agreement. It suggests that the diffusion kinetics are controlled by clustered rather than independent nanopores. This non-linearity of the diffusion characteristic time contrasts with previously used phenomenological models~\cite{Dupin2018}. Using expressions of $<T_n>$, one can also deduce the typical cluster sizes $\sqrt{n}a \approx \lambda <T_1> a$. Taking $R=75~\mu$m, $D=4.10^{-10}m^{2}/s$, and $a=1.4$~nm~\cite{song1996structure} gives sizes of clusters that range from about 10~nm ($n \approx 100$) at the lowest concentration to about 100~nm ($n \approx 10^4$) at the highest concentration. This clustering should also depend on membrane composition, which might explain the discrepancies in the reported $\alpha$HL pore concentrations between different lipid mixtures~\cite{Lemiere2013}.

\par We have used DIBs decorated with $\alpha$HL passive ion channels to mimic the passive molecular transport through biological cells. Using fluorescence imaging, we have quantified the effect of the pore concentration on the diffusion kinetics of a molecule from one cell to its neighbors. We have found that the diffusion kinetics are efficiently captured by a continuous time random walk model. We also found that the characteristic diffusion time scale varies as the inverse square root of the number of pores within the lipid bilayer, suggesting a pore clustering scenario. Such clustering has been evidenced numerically for transmembrane proteins~\cite{schmidt2008}. Taking into account this clustering for nanopores is thus likely to be relevant for molecular transport in real biological systems.

\begin{acknowledgments}
The authors deeply thank J.~Math\'e, J.-C.~Galas, A.~Estevez-Torres and A.~Senoussi for sharing experimental details, T.~Bertrand for fruitful theoretical discussions and acknowledge financial support from ANR (BOAT, ANR-17-CE30-0001).
\end{acknowledgments}

\bibliographystyle{apsrev4-1}
\bibliography{bibliography}
\end{document}